# Structural, magnetic and electric polarization properties of geometrically frustrated $YBaCo_4O_7$ and $DyBaCo_4O_7$ cobaltites

C. Dhanasekhar[1, 5(a)], A. K Das[1], A. Das [2, 3], S. K. Mishra [4], R. Rawat [4] and A. Venimadhav [5]

[1,5]Department of physics, Indian Institute of Technology, Kharagpur -721302, India

[2]Solid State Physics Division, Bhabha Atomic Research Centre, Mumbai 400085, India

[3]Homi Bhabha National Institute, Anushaktinagar, Mumbai 400094, India

[4]UGC-DAE Consortium for Scientific Research, Khandwa Road, Indore, 452017, Madhya Pradesh, India

[5]Cryogenic Engineering Centre, Indian Institute of Technology, Kharagpur -721302, India

**ABSTRACT –** In $RBaCo_4O_7$ (R = Ca, Y and Rare earth) cobaltite family, only $CaBaCo_4O_7$ shows 3D long range ferrimagnetic and spin driven electric polarization properties. In the present study, we have investigated the structural, magnetic and electric polarization properties in $YBaCo_4O_7$ and $DyBaCo_4O_7$ members of this family and the obtained results are compared with the $CaBaCo_4O_7$. The compound $YBaCo_4O_7$ showed a series of magnetic transitions in agreement with reported literature, additionally, a cluster glass behavior below 5 K is observed. Powder neutron diffraction studies on $DyBaCo_4O_7$ cobaltite showed an orthorhombic $Pbn2_1$ symmetry and signature of short-range $120^0$ magnetic correlations of kagome layer and a spin glass behavior below 65 K. Dielectric measurements on both the samples showed strong frequency dispersion at high temperature and frequency independent behavior at low temperature without dielectric anomaly. Pyroelectric current measurement has showed a broad peak around 50 K in both the samples; however, a careful analysis relates the peak to thermally stimulated depolarization current. This study signifies that the giant structural distortions and cobalt charge ordering at kagome layer are the key factors to drive both long range magnetic ordering and spin driven electric polarization in this cobaltite family.

(a) dsekhar21iitkgp@gmail.com

**INTRODUCTION.** – Cobalt oxides consistently showed much attention because of their rich variety of fascinating physical properties such as unconventional superconductivity [1], large thermoelectric power [2], spin state transitions [3], giant magnetoresistance [2,4], charge stripes [5]. In recent years a diverse interesting features, i.e. spin liquid phase [6] unidirectional THz absorption and magnetoelectric effects [7-8], and giant magnetic-field-induced polarization [9] were reported in $RBaCo_4O_7$ (R = Ca, Y and Rare earth) cobaltites. The crystal structure of these materials is built-up with alternating kagome and triangular layers, where the corners of the triangles occupied by magnetic cobalt ions and causes the frustration [10.11].

$CaBaCo_4O_7$ (CBCO) cobaltite belongs to the noncentrosymmetric orthorhombic structure (space group $Pbn2_1$) and was reported to undergo a 3D ferrimagnetic ordering below Tc = 70 K [10]. The ferrimagnetic ordering arises from the large structural distortion at kagome layers and strong spin exchange interactions between the triangular and kagome layer cobalt atoms along the c-axis. This compound shows a spin driven electric polarization below 70 K and is found to be pyroelectric in nature (electric polarization cannot switch with electric field) [9,12-14]. In our previous work, we have demonstrated switching from pyroelectric - ferrimagnetic ground state to ferroelectric - antiferromagnetic state in CBCO by replacing $Co^{2+}$ by $Ni^{2+}$ on kagome layer [15].

On the other hand, $YBaCo_4O_7$ (YBCO) also possesses noncentrosymmetric with hexagonal/trigonal structure (space group $P6_3mc$/ $P31c$) [11, 16-20]. With lowering temperature, a structural phase transition occurs initially at 310 K from the hexagonal to an orthorhombic phase $Pbn2_1$ and finally to a monoclinic structure $P2_1$ below 100 K [19]. The orthorhombic to monoclinic structural transition coincides with the onset of spin lattice coupling in this compound. This relieves partly the geometric frustration and drives it to an antiferromagnetically (AFM) ordered state below 100 K followed by a subsequent spin reorientation around 60 K [19-20]. Further, neutron diffraction studies show a strong spin-spin correlation between the triangular and kagome layer spins along the c axis, whereas the kagome layers (ab plane) spins shows a short-range $120^0$ correlations [16,17,20]. Room temperature X-ray diffraction measurements on the $DyBaCo_4O_7$ (DBCO) shows that the orthorhombic $Pbn2_1$ phase and magnetization measurements shows the Griffiths phase at low temperature [21].

To further understand the spin driven electric polarization properties in $RBaCo_4O_7$ cobaltites, it is essential to study the structural, magnetic and electric polarization properties of other members of this cobaltite family and compare their results. With this perspective, the $YBaCo_4O_7$ and $DyBaCo_4O_7$ members of this family were selected and studied their properties in systematic way. Although these materials have necessary symmetry conditions [19, 21] to have the spin driven electric polarization but one would expect a different origin of electric polarization as in $YBaCo_4O_7$, the $Y^{3+}$ (S=0) ion is nonmagnetic and $Dy^{3+}$ (S=5/2) has rare earth spin contribution. In $DyBaCo_4O_7$ one would also expect $4f$ - $3d$ interactions, which may further influence the magnetic ordering and electric polarization properties, whereas in the case of YBCO the cobalt sublattice only would expect to cause the spin driven electric polarization.

**EXPERIMENT DETAILS.** – The $DyBaCo_4O_{7+\delta}$ (DBCO) and $YBaCo_4O_{7+\delta}$ (YBCO) samples were synthesized by solid state reaction method following a procedure as reported previously [22-23]. Both powders of $YBaCo_4O_7$ and $DyBaCo_4O_7$ were obtained through a two-step solid state reaction method starting from stoichiometric high purity amounts of $Y_2O_3$, $Dy_2O_3$, $BaCO_3$, and $Co_3O_4$. The first heat treatment was made at 900 °C for 12 h. After regrinding the samples and pressing it into a pellet, the second sintering was done at 1100 °C for 24 h. After the final reaction the pellet was quenched to room temperature in order to stablize the 114 cobaltite phase [23]. The structural and magnetic properties of this cobaltite family is strongly influenced by the oxygen non stoichiometry. The oxygen stoichiometry of the as prepared samples were determined by idiometric titration [24], which shows the nearly stoichiometric ($\delta = 0.04$: for Y sample and $\delta = 0.07$ for the Dy sample) within the experimental limits. Phase purity of the samples is confirmed by x-ray powder diffraction at room temperature. The neutron diffraction measurements on DBCO samples were carried out on the PD2 powder neutron diffractometer ($\lambda=1.2443$ Å) at Dhruva reactor, Bhabha Atomic Research Centre, Mumbai, India. Rietveld refinement of the x-ray powder diffraction and neutron diffraction patterns were carried out using the FULLPROF program. DC and AC magnetic measurements were carried in a commercial VSM SQUID magnetometer (Quantum Design, USA). Semi adiabatic heat pulse technique is used to measure the specific heat in the temperature range of 2–150 K. Dielectric and pyroelectric current measurements were done in parallel plate capacitor geometry using Agilent 4291A and Keithley electrometer 6517A respectively. For dielectric and pyroelectric current ($I_p$) measurements pellets of thickness 0.5 mm and a diameter of 5 mm were used.

**RESULTS.** – The phase purity of the YBCO and DBCO samples at room temperature was verified by X-ray diffraction, where YBCO shows a single-phase hexagonal crystal structure with a P6$_3$mc space group and the DBCO sample showed orthorhombic Pbn2$_1$ symmetry, which are similar to the literature [11, 16, 21]. Neutron powder diffraction measurements on DBCO sample and Rietveld refinement of the NPD patterns recorded at 300 K and 6 K are shown in the figs.1 (a) & (b). The experimental data were refined using orthorhombic P$bn$2$_1$ symmetry and the obtained nuclear structure of DBCO at 6 K is shown in the fig.1(c). The neutron diffraction pattern on DBCO sample has showed no increase in the intensity of the fundamental Bragg reflections or superlattice reflections at 6 K indicating the absence of long range magnetic ordering. However, a broad hump is observed at Q ≈ 1.35 Å$^{-1}$ at 6 K and is shown in the inset of the fig.1 (b). This broad hump is in fact appears below 80 K is marked by an arrow in the fig.1 (d). A similar behavior was observed in the *YBCO and* other members of this family [11, 19, 25-27] and their occurrence was assigned to the short range magnetic correlations.

In case of YBCO samples, the short range correlations are followed by long range magnetic order [16,17,18], while in the case of DBCO the long range magnetic order is not established. The lattice parameters, unit cell volume and orthorhombic distortion (D) are given in **Table.1**. At 6 K the 'b' and 'c' lattice parameters are lower than at 300 K, while the 'a' lattice parameter is higher resulting overall a small decrease in unit cell volume.

In the present case, the orthorhombic distortion ($D = \left(\frac{b}{\sqrt{3}} - a\right)/a$) is nearly constant from room temperature (0.15 %) to the 6 K (0.16%) and is much smaller than that of CBCO (1.01% at 300 K and 1.82% at 6 K) [15]. YBCO shows multiple magnetic transitions with varying temperature [11,18]. The *M (T)* curves show an anomaly at 300 K which can be attributed to the structural phase transition ($T_{st}$) from hexagonal phase to the orthorhombic phase. As the temperature is further lowered, magnetization rises below 80 K and shows a maximum at 50 K and then drops down to 10 K, and finally a small increase is found below 10 K. The *M (T)* behaviour above 10 K matches with previous studies, however small increase below 10 K is not discussed. The isothermal magnetization (*M (H)*) measured at 5 K is shown in the inset of fig.2 (a). The linear nature of *M (H)* suggests the AFM nature of YBCO.

The frequency and temperature variation of in-phase ($\chi'(T)$) component of ac magnetization of YBCO is shown in the fig. 2(c). The $\chi'(T)$ curve shows a frequency independent peak at 80 K, which coincides with the sharp rise of dc magnetization, indicating the antiferromagnetic (AFM) transition. In addition, $\chi'(T)$ shows a broad hump at 50 K and again a sharp peak near 5 K. The peak position in $\chi'(T)$ at 5 K is strongly dependent on the measuring frequency and is shown in the inset of fig. 2 (b). Although the dc magnetization studies look similar to the previous studies but $\chi'(T)$ behaviour of the YBCO is different from the Valldor et al studies. In Valldor et al studies the $\chi'(T)$ shows a frequency dependent peak at 60 K and assigned to the spin-glass-like behaviour. The difference in the $\chi'(T)$ of both studies might be due to the disorder which arises from the changes in synthesis conditions [16, 28].

To understand the nature of this peak we have measured the Mydosh parameter [29]

$$\varphi = \frac{\Delta T_f}{T_f \Delta(\log_{10}(f))} \qquad (1)$$

which indicates the relative frequency shift in the peak temperature ($T_f$) of $\chi'(T)$ per frequency decade and is found to be ~ 0.094. In general, the $\varphi$ value is used to differentiate among various glassy systems; for spin glass $\varphi$ ~ 0.005 - 0.01, for cluster glass $\varphi$ ~ 0.03 - 0.06 and in case of superparamagnetic $\varphi$ > 0.1 [30]. The obtained $\varphi$ value in YBCO is higher than that of cluster glass, and this means that the interaction between the clusters is either very weak or completely absent. The frequency dependent freezing temperature of $\chi'(T)$ could not be fitted to the Arrhenius thermal activation mechanism. The shape of the *M (H)* at 5K is also linear, precluding the super paramagnetic (SPM) nature of the YBCO. The relaxation time ($\tau$) was analyzed using the expression for critical slowing down power law [29] given by

$$\tau = \tau_0 \left(\frac{T_f}{T_g} - 1\right)^{-zv} \qquad (2)$$

where $\tau_0$ is the microscopic spin relaxation time, $T_g$ is the glassy freezing temperature and *zv* denotes the critical exponent. The corresponding fitting is shown in the inset of **Fig. 2 (b)**, and the obtained values are; $\tau_0 = 3.45 \times 10^{-5}$ sec, $T_g = 3.7$ K and *zv* = 6 respectively. The large values of $\tau_0$ and *zv* indicate that YBCO has the freezing of magnetic clusters rather than the individual atomic spins.

The temperature variation of magnetization of DBCO sample measured under dc field ($H_{dc}$) of 0.05 T is shown in the **Fig. 3 (a).** With decreasing temperature, magnetization shows a peak at 70 K and this behavior is consistent with the literature [21], where it was assigned to an AFM ordering of the Co sublattice. However, our NPD study confirms the absence of the long range magnetic ordering. Further, below 70 K the magnetization increases due to the paramagnetic $Dy^{3+}$ ion contribution**.**

The *M (H)* curves measured at various temperatures below and above 70 K are shown in the fig. 3 (b)**.** At 5 K, *M (H)* shows a linear and reversible behavior below the critical field ($H_C$) of 2.2 T and above $H_C$, *M (H)* shows a linear behavior up to 7 T without saturation. A similar nonlinear behavior is also found at 30 K, 40 K and 50 K, with decreasing net *M* and finally linear *M (H)* behavior is observed at 90 K which matches with the *M (T)* behavior. Similar kind of *M (H)* and $H_C$ values were observed in other cobaltites in this family such as TbBaCo$_4$O$_7$ [31] and YbBaCo$_4$O$_7$ [32] while such a behavior is not observed in YBCO (inset of fig. 2 (a)), this indicates that the rare earth magnetic moment at the 'A' site is related to this field induced change in magnetization.

Further, to understand the magnetization of DBCO samples we have measured the temperature dependent in-phase ($\chi'$ (T)) and out of phase ($\chi''$ (T)) component of ac susceptibility under $H_{ac}$ ~ 10$^{-4}$ T. The obtained results are shown in the figs. 3(c) & (d). The $\chi'$ (T) shows a step like features in the vicinity of 70 K, which closely matches with M (T). Though it shows a frequency dependence, due to the increasing moment a meaningful analysis could not be carried. On the other hand, a clear frequency variation is found in the $\chi''$ (T). A finite value of $\chi''$ (T) and the corresponding shift in its peak position to higher temperature with increasing frequency suggesting a glassy nature. Experimentally, the freezing temperature ($T_f$) can be obtained from the maximum of $\chi'$ (T), or from the inflection point (or the maximum) of $\chi''$ (T) [30]. Both methods were extensively used to find the $T_f$ in spin glass (SG) systems [33-34]. Since the maximum of $\chi'$ (T), in our system is difficult to identify, we have used the maximum of $\chi''$ (T) to obtain the $T_f$. The φ were estimated from the $\chi''$ (T) peak and is found to be ~ 0.019, which falls within the spin glass range [29].

Further, the relaxation dynamics of the DBCO is examined using the power law given by eq. (2) and the results are shown in the inset of fig. 3 (d). From the fitting, the obtained values are $\tau_0$ =1.642x10$^{-13}$ sec, $T_a$=58.7 K and zv =10.2. These values are within the range of spin glass materials [29] and thus it may be concluded that DBCO is a spin glass. The absence of the long range magnetic ordering is further supported by the heat capacity measurement with absence of anomaly in the vicinity of 70 K and is shown in the inset of the fig. 1 (b). The heat capacity behavior matches with the previous studies where no significant anomaly is observed at magnetic ordering temperature [35]. The present ac magnetization and specific heat measurements complement the absence of long range magnetic ordering observed from the NPD.

Figures 4(a) &(c) shows the real part of dielectric permittivity (ε) as a function of temperature for different frequencies in YBCO and DBCO samples. Frequency independent ε was observed at low temperature and a step like increase with temperature is noticed for both the samples. Figures 4(b) & (d) shows the temperature dependent dielectric loss tangent (tan δ) as a function of frequency. The observed broad peak in tan δ shifted with increasing frequency to a higher temperature, indicating a thermally activated relaxation behavior. However, the ε shows no peak at the magnetic transition and indicates that there are no permanent electric diploes or the spin induced dipoles. The maximum peak position in the tan δ is fitted with the thermally activated Arrhenius mechanism (eq. (3)) and is shown in the inset of figs. 4(a) & (d) for the YBCO and DBCO samples.

$$\tau(T) = \tau_0 \exp\left(\frac{E_a}{K_B T}\right) \qquad (3)$$

where $E_a$ is the activation energy required for the relaxation process and $\tau_0$ is the pre-exponential factor. The $E_a$ and $\tau_0$ for the DBCO are 0.118(5) eV and 2.91X10$^{-11}$ s and 0.123(1) eV and 5.52X10$^{-11}$ s for YBCO respectively.

The pyroelectric current ($I_p$) is recorded while warming the samples, and the obtained results for the both samples are shown in the figs. 5 (a) & (b). In both samples, a broad $I_p$ peak near 45 K is observed, and the maximum peak temperature ($T_{max}$) depends strongly on the heating rate. The resulting electric polarization ($P$) for the corresponding $I_p$ is shown in the figs. 5 (c) & (d) which confirm that the polarizations and their saturation value depend on heating rate. This behaviour is in contrast with CBCO, which shows a sharp λ like $I_p$ peak in the vicinity of the 65K and the peak position is found to be independent of heating rate [15]. The broad $I_p$ peak and its dependence on the heating rate suggest the extrinsic origin of the $P$. This $I_p$ may be due to the thermally stimulated free charge carriers (TSFCC) observed in several spin driven ferroelectric materials [36-37]. Further, the extrinsic $I_p$ can be understood from the Bucci-Fieschi-Guidi framework [37-38] where the relation between heating rate ($b$) and the maximum peak temperature ($T_m$) of $I_p$ is given

$$\ln \frac{T_m^2}{b} = \frac{E_a}{k_B T_m} + \ln \frac{\tau_0 E_a}{k_B} \qquad (4)$$

where $E_a$ is the activation energy, $\tau_0$ is the relaxation time and b is the heating rate. Insets of fig. 5 (c) & (d) show the plot of ln ($T^2_m$ /b) versus $1/T_m$ for the DBCO and YBCO samples. The obtained $E_a$ and $\tau_0$ values for the DBCO and YBCO samples are $E_a$ = 0.0561 eV; $\tau_0$ = 9.19X10$^{-7}$ s and $E_a$ = 0.064 eV; $\tau_0$ = 7.35X10$^{-8}$ s respectively. The sign of TSFCC is determined by the type of the trapped charges during the $E_p$. In the present case, the TSFCC probably originate from the trapped holes, therefore it has the same sign as the $E_p$, as shown in the figs. 5 (a) & (b), which is similar to the DyMnO$_3$ [39].

**DISCUSSION AND CONCLUSIONS.** – YBCO and DBCO samples belong to the noncentrosymmetric space group similar to the CBCO but the former samples show the absence of spin driven electric polarization. The main contrast is in magnetic ground state, which is strongly associated with the strength of orthorhombic distortion (D). At low temperature, the orthorhombic distortion (D in %) is 0.16, 0.34 and 1.82 for the DBCO, YBCO [19] and CBCO [15] samples respectively. The high D values in the CBCO lefts the inherent magnetic frustration in the ab plane (or kagome layer) and drives the long-range 3D ferrimagnetic ordering below 70 K. On contrary the low D values in the DBCO and YBCO samples suggest that the magnetic frustration in the kagome layers remain unchanged down to low temperature and causes the short-range magnetic ordering and glassy behaviour in the both samples. Apart from giant structural distortions, the other major difference between the CBCO and the YBCO and DBCO might be the cobalt charge ordering in the kagome layers. In CBCO, it is showed that the kagome layer contains the mixed valances of Co$^{2+}$, Co$^{3+}$ or Co$^{2+}$(L) and causes the spin driven electric polarization [10,40]. On contrary, according to the nominal composition of the YBCO and DBCO cobaltites, the kagome layers might contains the Co$^{2+}$ valance state, which states the absence of the cobalt charge ordering at kagome layers in both samples.

In conclusions, this study clearly shows a close correlation between the orthorhombic distortion, cobalt charge ordering and spin driven electric polarization in this RBaCo$_4$O$_7$ (R = Ca, Y and Rare earth) cobaltite family. The magnetization studies on YBaCo$_4$O$_7$ show the structural transition ($T_{st}$ =300 K), AFM ordering ($T_N$ = 80 K) and cluster glass behavior ($T_g$ = 3.6 K). NPD studies on DyBaCo$_4$O$_7$ show evidence for the short range correlations below 70 K and ac magnetization studies show spin glass behavior. Both samples show the absence of spin induced ferroelectric polarization and extrinsic pyroelectric peaks with the thermally stimulated depolarization currents. This study further indicates a similar kind of extrinsic electric polarization properties would be expected from other rare earth members of this family.

***

The authors of IIT Kharagpur acknowledge DST, India for FIST project and IIT Kharagpur funded VSM SQUID magnetometer. We thank Ripandeep Singh, Solid State Physics Division, Bhabha Atomic Research Centre, Mumbai, for his assistance in recording the neutron diffraction data.

**Table 1:** The refined unit cell parameters, lattice volume, orthorhombic distortion (D in %) and reliability factors of DyBaCo$_4$O$_7$ obtained from Rietveld refinement of the neutron diffraction at 300 K and at 6 K. Orthorhombic crystal structure, space group of P$bn2_1$.

| Parameters | 300 K | 6 K |
|---|---|---|
| a (Å) | 6.2959(12) | 6.2949(12) |
| b (Å) | 10.9200(22) | 10.9217(21) |
| c (Å) | 10.2415(8) | 10.2158(7) |
| V (Å$^3$) | 704.120(0.212) | 702.365(0.201) |
| Distortion (D) % | 0.153 | 0.164 |
| R$_{Bragg}$ (%) | 4.26 | 6.08 |
| R$_f$ (%) | 2.65 | 3.93 |
| $\chi^2$ | 2.35 | 2.99 |

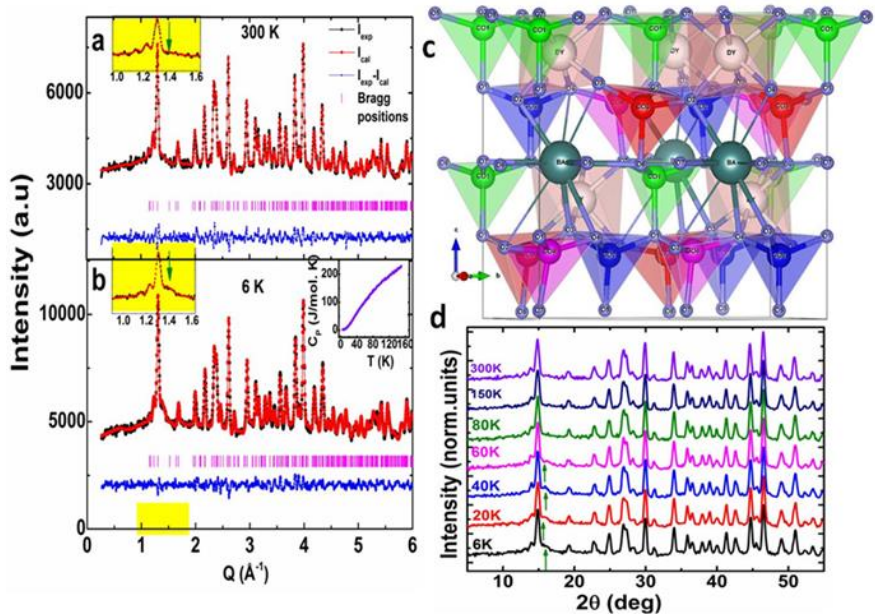

**Fig. 1:** (Color online) Room temperature (a) and low temperature (b) experimental NPD data of the DyBaCo$_4$O$_7$ (black dots) with the simulated curve (red lines). Here, the blue lines show the difference between the experimental and simulated curve and vertical magenta lines show the corresponding nuclear Bragg positions. The top insets of (a) & (b) show the neutron diffraction patterns in a selected range. The right inset of the (b) shows the temperature variation of specific heat. The obtained nuclear structure at 6 K from NPD is shown in (c) and (d) shows NPD measured at various temperatures.

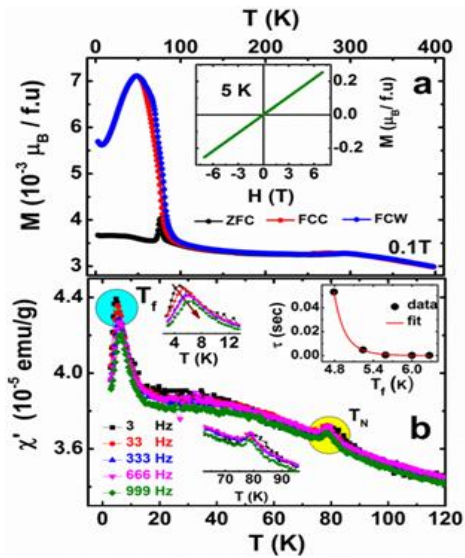

**Fig. 2:** (Color online). (a) $M(T)$ curves of YBaCo$_4$O$_7$ measured under H$_{dc}$ of 0.1 T and (b) shows the $\chi'(T)$ measured under H$_{ac}$ ~10$^{-4}$ T at different frequencies. Inset of (a) illustrates the $M(H)$ curves at 5 K. The right inset of (b) shows the power law fit (to eq. 1.) of low temperature peak around 5 K.

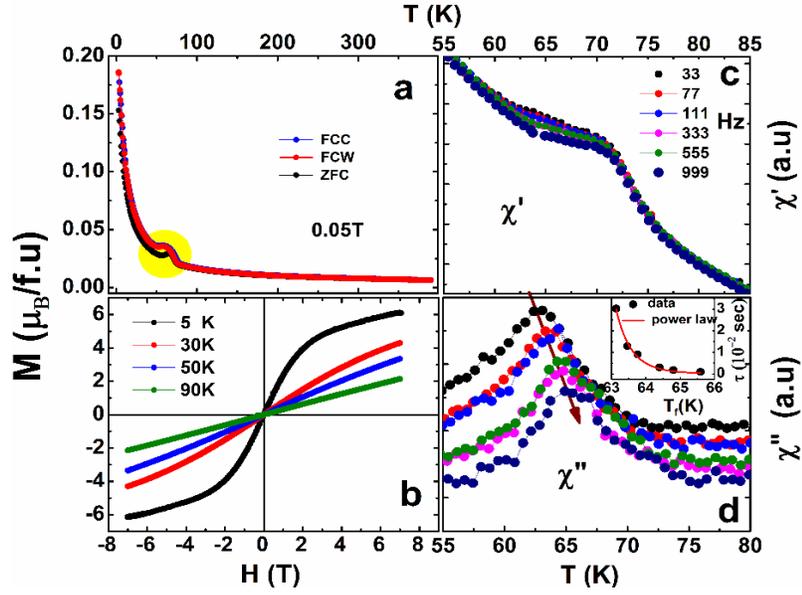

**Fig. 3:** (Color online). *M (T)* curves of DyBaCo$_4$O$_7$ measured under a magnetic field of 0.05 T (a) and *M (H)* curves recorded at various temperatures (b). The (c) & (d) shows χ' (*T*) & χ" (*T*) measured under H$_{ac}$ ~10$^{-4}$ T at different frequencies. The inset of inset (d) shows the power law fit (eq. 2.) of maximum χ" (*T*) peak temperature.

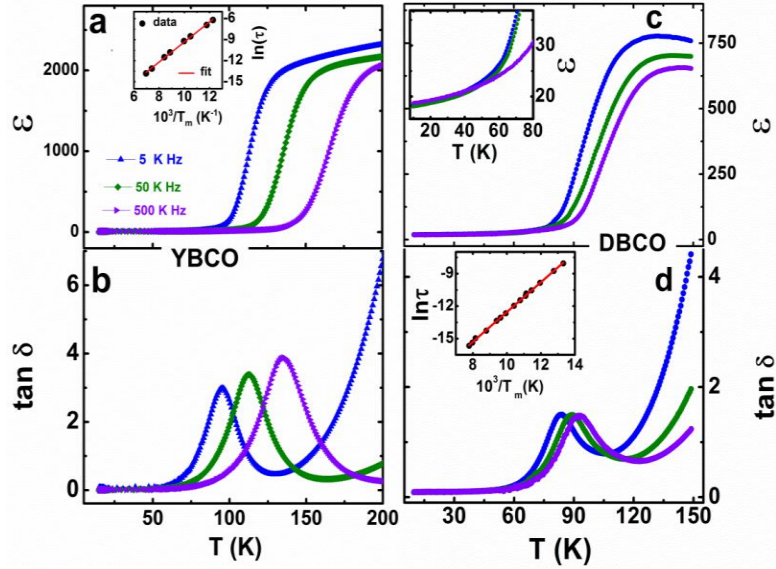

**Fig. 4:** (Color online). The (a) & (c) shows ε(T) curves for the YBaCo$_4$O$_7$ and DBCO samples and corresponding tan δ of these samples are shown in the (b) & (d). The insets of (a) & (d) show the Arrhenius fit (**Eq. 3.**) for the maximum peak temperatrure of tan δ.

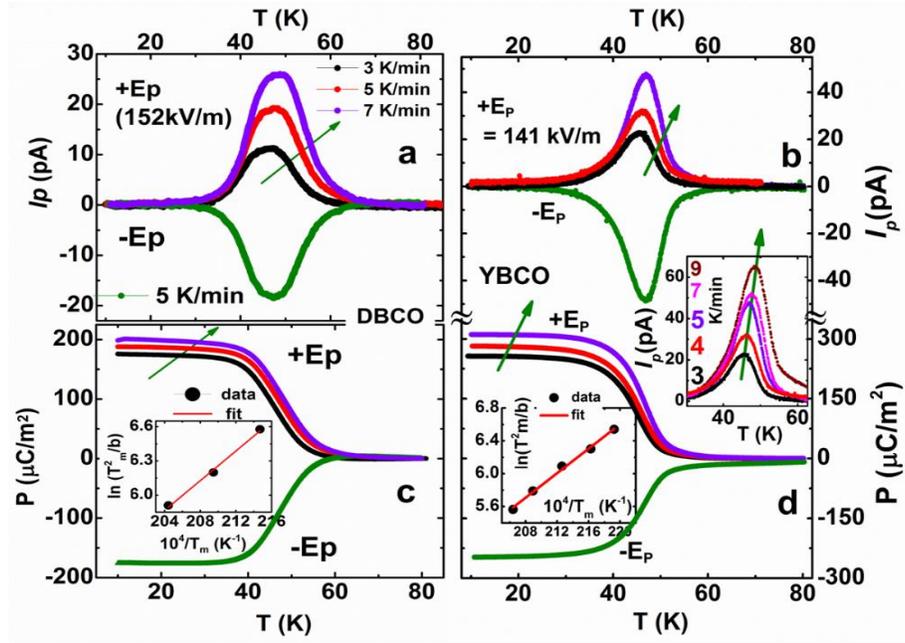

**Fig. 5:** (Color online). The (a) & (c) show temperature variation of $I_p$ and corresponding polarization for DyBaCo$_4$O$_7$ measured for $E_p = \pm 152$ kVm$^{-1}$ under various heating rates. The (b) & (d) show $I_p$ versus temperature and corresponding polarization for YBaCo$_4$O$_7$ measured under $E_p = \pm 141$ kVm$^{-1}$. The inset (c) & (d) shows the ln ($T^2_m$/b) vs. 1/$T_m$ fitting and is described in the main text.